\documentstyle[preprint,aps,eqsecnum]{revtex}
\begin{document}


\title{THE COEFFICIENTS OF FRACTIONAL PARENTAGE OF NUCLEAR SHELL MODEL}

\author{A.Deveikis, G.Kamuntavi{\v c}ius}

\address{Vytautas Magnus University, Kaunas 3000, Lithuania}

\date{\today}

\maketitle

\begin{abstract}
     A new procedure has been developed for the calculation of shell model
 coefficients of fractional parentage. This procedure is free from the
 numerical diagonalization, orthogonalization and even group theoretical
 antisymmetric states classification.
\end{abstract}

\columnsep=8 mm

\section{INTRODUCTION}

New phenomena, obtained recently in nuclear physics (neutron - rich
nuclei and neutron halo) cannot be described in shell- model approximation
and require high - quality wave function. One of the most efficient methods
for the construction of many - particle wave functions with well - defined
permutational symmetry and exact quantum numbers ($J$ - total angular
momentum and $T$  - isospin) is the iterative procedure for the calculation
of coefficients of fractional parentage (CFP) developed in classical works
\cite{1,2}. However, due to the high degree of degeneracy of shell - model
states, the set of exact quantum numbers in many cases is not rich enough
to classify different antisymmetrical states of system.

The first method for classifying antisymmetrical states  \cite{3}
involves the construction and diagonalization of the matrices representing
the quadratic Casimir operators for the appropriate special unitary,
orthogonal or symplectic groups. The real difficulty in such a case is
finding the Casimir operator and knowing its eigenvalues.

For many applications the precise specification of the states is not
required. It is merely sufficient that the states be orthogonal. The second
method \cite{4} uses the Redmond iteration technique in conjunction with
numerical Schmidt orthogonalization or Grammian matrix diagonalization.
Unfortunately, although this procedure can  generate complete sets for
large spaces, the buildup of numerical errors in the process of imposing
orthonormality within the overcomplete set of antisymmetrical states
causes many problems.

We shall here present a new and very simple method of CFP calculation,
based on complete rejection of group - theoretical antisymmetrical shell
model states classification and even of numerical diagonalization or
orthogonalization.

\section{DEFINITIONS AND NOTATIONS}

Coefficients of fractional parentage are defined as the coefficients
for the expansion of antisymmetrical wave - function in terms of the complete
set of the vector - coupled parent states with a lower degree
of antisymmetry. In the simplest case (one particle CFP) this expansion is:

\begin{eqnarray*}
\left\langle x_1,\ldots,x_{N-1},x_N |
    K \alpha J T M_J M_T
\right\rangle =
\displaystyle{
    \sum_{  \stackrel{\bar{K} \bar{\alpha} \bar{J} \bar{T}}
		     {\varepsilon_N \ell_N j_N}
	 }
}
\left\langle
    x_1,\ldots,x_{N-1};x_N |
    \left(\bar{K} \bar{\alpha} \bar{J} \bar{T};
	  \varepsilon_N \ell_N j_N
    \right)
    J T M_J M_T
\right\rangle
\times \\
\times
\left\langle
    \bar{K} \bar{\alpha} \bar{J} \bar{T}; \varepsilon_N \ell_N j_N
    \parallel
    K \alpha J T
\right\rangle
\ .
\end{eqnarray*}

Here one -- particle variables are $x_i\equiv{\bf r}_i\sigma_i\tau_i$
(set of corresponding radius -- vector, spin and isospin variables)
and quantum numbers $\varepsilon_i \ell_i j_i$
(oscillator energy, orbital momentum and total momentum).

The corresponding set of many - particle state quantum numbers are
$K$ -- configuration (it can consist of only one shell nucleons),
$\alpha$ -- all necessary additional quantum numbers and
$JT$  -- exact quantum numbers. Semicolon
means that the corresponding wave - function is fully antisymmetric
with respect to interchanges among the nucleons variables, distributed
in different sides of this sign. Notice the use of parentheses to represent
angular momentum coupling.
$\langle\ldots;\ldots\parallel\ldots\rangle$ is the coefficient of
the fractional parentage identificator.

The expression given above is written down in $j - j$ coupling, but
final results are not dependent on the coupling scheme.

To calculate coefficients of fractional parentage it is enough to
calculate $N$ - particle antisymmetrization operator matrix into the basis of
right - hand functions (with a lower degree of antisymmetry). As usual, one
assumes that all $(N-1)$ particle CFPs have been obtained at a previous
stage. The antisymmetrizer
\begin{eqnarray*}
A
\equiv
A_{1,\ldots,N}
=
{1\over{N!}}
\sum_{P\in S_N}
\delta_P P
\end{eqnarray*}
is normalized such that $A^2=A$ and could be presented as
\begin{eqnarray*}
A_{1,\ldots,N}
=
A_{1,\ldots,N-1}
Y
A_{1,\ldots,N-1}
\ ,
\end{eqnarray*}
where
\begin{eqnarray*}
Y
=
{1\over N}
\left\{
	1-(N-1)P_{N-1,N}
\right\}
\ .
\end{eqnarray*}
Here the operator $P_{N-1,N}$ simply interchanges the coordinates
$x_{N-1}$ and $x_{N}$.
Because the above mentioned basis is antisymmetrical by
$x_1,\ldots,x_{N-1}$ variables permutation,
matrix ${\bf A}$ equals the matrix of a far simpler
operator  $Y$:
\begin{eqnarray*}
\lefteqn{
\left\langle
\left(
\bar{K} \bar{\alpha} \bar{J} \bar{T},
\varepsilon_N \ell_N j_N
\right)
    J T M_J M_T
|Y|
\left(
\bar{K'} \bar{\alpha'} \bar{J'} \bar{T'},
\varepsilon'_N \ell'_N j'_N
\right)
J T M_J M_T
\right\rangle
}
\\
\\
&
=
{1\over N}
\left\{
        \delta_{\bar{K},\bar{K'}}
        \delta_{\bar{\alpha},\bar{\alpha'}}
        \delta_{\bar{J},\bar{J'}}
        \delta_{\bar{T},\bar{T'}}
        \delta_{\varepsilon_N,\varepsilon'_N}
        \delta_{\ell_N,\ell'_N}
        \delta_{j_N,j'_N}
\right.
\\
\\
&
\left.
+
(-1)^{j_N + j'_{N} + \bar{J} + \bar{T} + \bar{J'} + \bar{T'}}
(N-1)
[(2 \bar{J} + 1)(2 \bar{T} + 1) (2 \bar{J'} + 1) (2 \bar{T'} + 1)]
^{1\over 2}
\right.
\\
\\
&
\left.
\times
\displaystyle{
  \sum_{\tilde{K}\tilde{\alpha} \tilde{J} \tilde{T}}
}
\left\{
  \matrix{
           j'_N \tilde{J} \bar{J}      \cr
           j_N J \bar{J'}              \cr
         }
\right\}
\left\{
  \matrix{
           {1\over2} \tilde{T} \bar{T} \cr
           {1\over2} T \bar {T'}       \cr
         }
\right\}
\left\langle
 \tilde{K} \tilde{\alpha} \tilde{J} \tilde{T};
 \varepsilon'_N \ell'_N j'_N
 \parallel
   \bar{K} \bar{\alpha} \bar{J} \bar{T}
 \right\rangle
\right.
\\
\\
&
\times
\left.
\left\langle
    \tilde{K} \tilde{\alpha} \tilde{J} \tilde{T};
 \varepsilon_N \ell_N j_N
 \parallel
    \bar{K'} \bar{\alpha'} \bar{J'} \bar{T'}
\right\rangle
\right\}
\ .
\end{eqnarray*}
{\bf A} is the symmetrical real matrix -- projection
$\left( {\bf A}^{+} = {\bf A} = {\bf A}^{*}\ ,\
{\bf A}{\bf A}={\bf A}\right).$
Its eigenvalues are only zeros and units,
and the spectral decomposition \cite{5} is
\begin{eqnarray*}
{\bf A}_{n\times n}
=
\tilde{\bf F}_{n\times r}
\left(
     \tilde{\bf F}^{+}
\right)_{r\times n}
\ .
\end{eqnarray*}

     The subscripts indicate the dimension  of the corresponding matrix.
$n$ equals the dimension of the basis,  $r$ -- rank of matrix  ${\bf A}$.
Every column of  $\tilde{\bf F}$
is eigenvector of matrix  ${\bf A}$, corresponding to a unit eigenvalue.
The  normalization condition of eigenvectors is
\begin{eqnarray*}
\left(
    \tilde{\bf F}^{+}
\right)_{r\times n}
\cdot
\tilde{\bf F}_{n\times r}
=
{\bf 1}_{r\times r}
\ .
\end{eqnarray*}

The matrix elements of $\tilde{\bf F}$ are coefficients
of fractional parentage \cite{6}.

\section{NEW RESULTS}

    Condition ${\bf A}_{n\times n}{\bf A}_{n\times n}={\bf A}_{n\times n}$
means that every  ${\bf A}_{n\times n}$ column is its
eigenvector with unit eigenvalue. However, they are not normalized and even
orthogonal, and only  $r$       of them are linearly independent. The simplest
method of CFP matrix calculation is based on the observation that the spectral
decomposition of ${\bf A}$     is not defined  uniquely. The
possibility exists of a free choice of orthogonal matrix  ${\bf G}_{r\times r}$
defined as
\begin{eqnarray*}
{\bf A}
=
\tilde{\bf F}{\bf G}{\bf G}^{+}\tilde{\bf F}^{+}
\equiv
{\bf F}{\bf F}^{+}
\ ,
\end{eqnarray*}
because
\begin{eqnarray*}
{\bf F}^{+}{\bf F}
\equiv
{\bf G}^{+}
\tilde{\bf F}^{+}
\tilde{\bf F}
{\bf G}
=
{\bf G}^{+}
{\bf G}
=
{\bf 1}
\ .
\end{eqnarray*}

     Orthogonal matrix has  $r(r-1)/2$	 free parameters, so we
can choose it in a way which allows us to fix the corresponding number of
{\bf F} elements. The best choice is
\begin{eqnarray*}
F_{ij} = 0\,
\qquad
\hbox{\rm if}
\quad
1\leq i < j \leq r
\ .
\end{eqnarray*}

This means that the upper triangle of matrix ${\bf F}$  equals zero.
In such a case we can obtain the solution of matrix equation
${\bf A}={\bf F}{\bf F}^{+}$
in the following way. Starting with the well -- known Redmond result \cite{2}
for the first row
\begin{eqnarray*}
F^2_{11} = A_{11} \ ,
\quad
F_{j1} = A_{1j}/F_{11}
 \ ,
\end{eqnarray*}
 we can present the following columns of ${\bf F}$ in the form:
\begin{eqnarray*}
F^2_{ii} = A_{ii} - \sum_{k=1}^{i-1} F_{ik}^2 \ , \\
F_{ji} = {1\over F_{ii}}
	\left\{
	     A_{ij} - \sum_{k=1}^{i-1} F_{ik}F_{jk}
	\right\}
\end{eqnarray*}
for every value of $i = 2,3, ..., r$ and the corresponding set of
$j = i+1,i+2, ..., n$.
Positive values of $F_{ii}$  are convenient, because the
overall sign of CFP vector is arbitrary. Obviously, if ${\bf F}$ is
constructed in this way, it  fulfils the condition
${\bf F}^{+}{\bf F}={\bf 1}$.

So to obtain the spectral decomposition
of an antisymmetrization operator
matrix with rank $r$, it is enough to calculate only the $r$
linearly independent rows of this matrix.
The calculation begins with the first matrix  ${\bf A}$ row. If
\begin{eqnarray*}
A_{11} = 0 \ ,
\end{eqnarray*}
then every element of the first row and the first column equals zero,
because by definition
\begin{eqnarray*}
A_{11} =
\sum_{j=1}^n A^2_{1j}
\ .
\end{eqnarray*}
This means that the corresponding basic function is not presented
in fractional parentage expansion (it may be,
that it has some hidden symmetry, wrong quantum numbers, etc.).
Such a function has to be omitted from all the expansions
in every case, when  $A_{ii}=0$ appears. Because matrix  ${\bf A}$
is symmetrical, every next row calculation  starts from diagonal element
$(A_{ll})$.
Then it is necessary to find the corresponding minor
$({\bf A}_{l\times l})$
determinant value. If it equals  zero, $l$--th row is
linearly dependent on $(l-1)$ the just calculated ones, because this
determinant equals calculated rows Grammian:
\begin{eqnarray*}
A_{ij} =
\sum_{k=1}^n A_{ik} A_{jk}
\ .
\end{eqnarray*}

Only in the case when $A_{ll}\not = 0$ and $det\,{\bf A}_{l\times l}
\not = 0$, is it worth finishing the $l$--th row calculation.
Because  ${\bf A}$ is a projection by definition,
the necessary number ($n$) of basic states is included
only in the case when the condition is fulfilled:
\begin{eqnarray*}
\sum_{j=1}^n A^2_{lj}
=
A_{ll}
\ .
\end{eqnarray*}

It is also useful to know the sum of matrix   ${\bf A}$
diagonal elements, because
\begin{eqnarray*}
Sp\,{\bf A} = r
\ .
\end{eqnarray*}

The expansions which have been presented allow after the calculation
of every matrix element of
${\bf A}$ ($A_{ij}$)
to obtain the corresponding CFP ($F_{ji}$).

So the calculation of one linearly independent row of ${\bf A}$
gives us the coefficients for one antisymmetrical state expansion
(one column of CFP matrix).
Columns of ${\bf F}$ can be numbered by the positive integer numbers
$\alpha = 1,2,\ldots,r$. In this  definition, $\alpha$ also equals
the number of zeros present in the corresponding column plus one.

\section{COMPUTATION RESULTS}

The general formalism described above was implemented in computer code to
generate complete sets of CFP for the $j = 1/2, 3/2, 5/2$ shells in isospin
formalism. Angular momentum coupling and antisymmetrization are orthogonal
transformations, so is possible precise arithmetic with quantities no more
sophisticated as square root of integer number. The CFP computational
procedure required four different types of arithmetics:
\begin{itemize}
\item algebraic and relation operations between fractions composed of
  16 - bit integer numbers for operating with quantities like
  angular momentum.
\item square root of 64 - bit integer numbers.
\item algebraic and relation operations between numbers
  represented in the form $n/m/\sqrt{k}$, where n,m,k are 64 - bit
  integer numbers for both middle scale and 6j coefficient calculations.
\item algebraic and relation operations between numbers represented in
  the form $n/m/\sqrt{k}$, where n,m,k are arbitrary length integer
  numbers accomplished as variable long vector with prescribed
  position weight for large scale calculations in evaluation ${\bf A}$ and
  ${\bf F}$ matrices. Orbit $j = 5/2$ was calculated
  with 20 - position vectors and position weight equal to 1000.
\end{itemize}

As a preliminary step complete enumeration of the states in isospin formalism
was carried out. By means of combinatorial calculations  all JT and number of
states with the same JT were found. This enable find out all permited CFP and
to characterized them uniquely by means of parent and daugther states.
Results for the $j = 5/2$ nuclear shell are presented in the table.

\bigskip

\begin{tabular}{|c|c|rrrrrrr|} \hline
N  &  T  &  & & & n(J)m  & & & \\ \hline
 2  &  0 & 1(1)1   & 1(3)1     & 1(5)1    &           &           &           &           \\
      &  1 & 1(0)1   & 1(2)1     & 1(4)1    &           &           &           &           \\ \hline
 3  &  1/2 & 2(1/2)1   & 4(3/2)1     & 6(5/2)2    & 5(7/2)2   &
4(9/2)1   & 3(11/2)1  & 2(13/2)1  \\
      &  3/2 & 2(3/2)1   & 3(5/2)1     & 2(9/2)1    &           &           &           &           \\ \hline
 4  &  0 & 2(0)2   & 7(2)3     & 8(3)1    & 8(4)3   & 7(5)1   & 5(6)2   & 2(8)1   \\
      &  1 & 7(1)2   & 10(2)2    & 11(3)3   & 11(4)2  & 9(5)2   & 6(6)1   & 4(7)1   \\
      &  2 & 1(0)1   & 3(2)1     & 3(4)1    &           &           &           &           \\ \hline
 5  &  1/2 & 9(1/2)2   & 16(3/2)3    & 21(5/2)4   & 22(7/2)4  & 21(9/2)4  & 17(11/2)3 & 13(13/2)2 \\
      &        & 8(15/2)1  & 5(17/2)1    &            &           &           &           &           \\
      &  3/2 & 6(1/2)1   & 11(3/2)2    & 14(5/2)2   & 14(7/2)2  & 13(9/2)2  & 10(11/2)1 & 7(13/2)1  \\
      &  5/2 & 3(5/2)1   &             &            &           &           &           &           \\ \hline
 6  &  0 & 11(1)3  & 17(2)1    & 20(3)5   & 20(4)2  & 18(5)3  & 15(6)2  & 11(7)2  \\
      &        & 4(9)1   &             &            &           &           &           &           \\
      &  1 & 6(0)2   & 17(1)1    & 26(2)5   & 30(3)3  & 30(4)5  & 26(5)2  & 21(6)3  \\
      &        & 15(7)1  & 9(8)1     &            &           &           &           &           \\
      &  2 & 7(1)1   & 10(2)1    & 11(3)1   & 11(4)1  & 9(5)1   &           &           \\
      &  3 & 1(0)1   &             &            &           &           &           &           \\ \hline
\end{tabular}

\begin{tabular}{|c|c|rrrrrrr|} \hline
 7  &  1/2 & 14(1/2)2  & 25(3/2)3    & 32(5/2)4   & 35(7/2)4  & 34(9/2)4  & 29(11/2)3 & 22(13/2)2 \\
      &        & 15(15/2)1 & 10(17/2)1   &            &           &           &           &           \\
      &  3/2 & 10(1/2)1  & 18(3/2)2    & 23(5/2)2   & 24(7/2)2  & 23(9/2)2  & 18(11/2)1 & 14(13/2)1 \\
      &  5/2 & 6(5/2)1   &             &            &           &           &           &           \\ \hline
 8  &  0 & 4(0)2   & 17(2)3    & 20(3)1   & 20(4)3  & 18(5)1  & 15(6)2  & 7(8)1   \\
      &  1 & 17(1)2  & 26(2)2    & 30(3)3   & 30(4)2  & 26(5)2  & 21(6)1  & 15(7)1  \\
      &  2 & 3(0)1   & 10(2)1    & 11(4)1   &           &           &           &           \\ \hline
 9  &  1/2 & 9(1/2)1   & 16(3/2)1    & 21(5/2)2   & 22(7/2)2  & 21(9/2)1  & 17(11/2)1 & 13(13/2)1 \\
      &  3/2 & 11(3/2)1  & 14(5/2)1    & 13(9/2)1   &           &           &           &           \\ \hline
 10 &  0 & 5(1)1   & 8(3)1     & 7(5)1    &           &           &           &           \\
      &  1 & 3(0)1   & 10(2)1    & 11(4)1   &           &           &           &           \\ \hline
 11 &  1/2 & 6(5/2)1   &             &            &           &           &           &           \\ \hline
 12 &  0 & 1(0)1   &             &            &           &           &           &           \\ \hline
\end{tabular}

\bigskip
First column contains nucleon number N in the shell, second stands for
total isospin T, last column gives total J and two additional numbers:
n - CFP number for specified JT (the dimension of corresponding A matrix);
m - stands for number of states with the same JT (the rank of a
corresponding A matrix). Thus we can have $n \times m$ CFP with the same JT.
Though generally $j = 5/2$ nuclear shell has 3359 CFP, computational method
determines $m(m-1)/2$ zeros by zeroing ${\bf F}$ matrix upper right  corner.
This gives 141 zeros in complete CFP set. Actual number of CFP equal to zero
is 328 due possible nonincounted symmetries.

To illustrate the evaluation of the CFP consider N=4 state with J = 6, T = 0.
As follows from the combinatorial calculation in this case  ${\bf A}$ matrix
dimension is 5 and rank 2.

\[ {\bf A}^{[6,0]}_{4} = \left( \begin{array}{rrrrr}
\frac{5}{12} & \frac{7}{12 \sqrt{33}} & \frac{14}{3 \sqrt{462}}  & \frac{-21}{2 \sqrt{2002}}  & \frac{28}{ \sqrt{6006}}  \\
             & \frac{53}{396}         & \frac{-56}{99 \sqrt{14}} & \frac{-147}{22 \sqrt{546}} & \frac{14}{33 \sqrt{182}} \\
             &                        & \frac{49}{99}            & \frac{21}{11 \sqrt{39}}    & \frac{35}{33 \sqrt{13}}  \\
             &                        &                          & \frac{175}{286}            & \frac{-21}{143 \sqrt{3}} \\
             &                        &                          &                            & \frac{49}{143}
          \end{array} \right) \]

Computed matrix is matrix projection, its rows and columns are labeled
by CFP left - hand bracket indices. ${\bf A}$ matrix spectral decomposition
gives ${\bf F}$ matrix, which has 2 columns(as predicted by combinatorial
calculation) and contains CFP as matrix elements. To facilitate comprehension
${\bf F}$ matrix is displayed in one column.

\begin{eqnarray*}
 \langle 3(2): \; \: 7/2, 1/2; 5/2, 1/2||4(2): 6, 0 \rangle & = & \frac{5}{2 \sqrt{15}}    \\
 \langle 3(1): \; \: 7/2, 1/2; 5/2, 1/2||4(2): 6, 0 \rangle & = & \frac{7}{6 \sqrt{55}}    \\
 \langle 3(1): \; \: 9/2, 1/2; 5/2, 1/2||4(2): 6, 0 \rangle & = & \frac{28}{3 \sqrt{770}}  \\
 \langle 3(1): 11/2, 1/2; 5/2, 1/2||4(2): 6, 0      \rangle & = & \frac{-63}{\sqrt{30030}} \\
 \langle 3(1): 13/2, 1/2; 5/2, 1/2||4(2): 6, 0      \rangle & = & \frac{56}{\sqrt{10010}}  \\
 \langle 3(2): \; \: 7/2, 1/2; 5/2, 1/2||4(1): 6, 0 \rangle & = & 0                         \\
 \langle 3(1): \; \: 7/2, 1/2; 5/2, 1/2||4(1): 6, 0 \rangle & = & \frac{6}{\sqrt{330}}      \\
 \langle 3(1): \; \: 9/2, 1/2; 5/2, 1/2||4(1): 6, 0 \rangle & = & \frac{-21}{\sqrt{1155}}   \\
 \langle 3(1): 11/2, 1/2; 5/2, 1/2||4(1): 6, 0      \rangle & = & \frac{-49}{\sqrt{5005}}   \\
 \langle 3(1): 13/2, 1/2; 5/2, 1/2||4(1): 6, 0      \rangle & = & \frac{-21}{\sqrt{15015}}
\end{eqnarray*}

\bigskip
CFP are denoted
$ \langle \bar{N} (\bar{\alpha}): \bar{J}, \bar{T}; j,t||N(\alpha): J,T
\rangle $.
First number in the bracket is $N-1$, second number closed in parentheses
stands for parent state number with the same total angular momentum $\bar{J}$
and isospin $\bar{T}$, shell's j and t stands after semicolon.
The same is for daughter state in the right - hand. Columns of matrix
${\bf F}$ form orthonormalized sets of CFP: columns with the same $\alpha$
are normalized to unit and orthogonal.

\section{CONCLUSIONS}

The suggested method of CFP calculation is not dependent on a coupling
scheme and configuration complexity. It is the same in every case.
So, it is useful in the description of highly excited states,
when we try to investigate more interesting phenomena.
The method is useful for intrinsic CFP calculation, when the system
centrum of mass state is under control in every antisymmetrical basic
function. It is for this reason that we have introduced the original CFP
abbreviation and defined the procedure for arbitrary configuration.
How about possible numerical instabilities, we have developed
arithmetic, which allows calculations without approximations. Numbers
are represented in the form: $n/m/\sqrt{k}$, where n,m,k are integer numbers,
and all calculations are made with numbers in this form. Our computational
experience shows that method has high accuracy, therefore real numbers
are insufficient. We have used 25 Mhz AT - 386 for calculation of CFP of
the nuclear shell j = 5/2 and it take us about 20 hours.

\vskip 1 true cm

This work was supported, in part, by a Soros Foundation Grant
awarded by the American Physical Society.


\end{document}